\documentclass[]{spie}  %>>> use for US letter paper
\usepackage[]{graphicx}
\usepackage{amsmath,amssymb,amsfonts}
\usepackage{algorithmic}
\usepackage{graphicx}
\usepackage{textcomp}

\usepackage{hyperref}

\title{Automated Detection of Microaneurysms in Color Fundus Images using Deep Learning with Different Preprocessing Approaches } 

\author{Meysam Tavakoli\supit{a} Sina Jazani\supit{b} and Mahdieh Nazar\supit{c}
\skiplinehalf
\supit{a}Department of Physics, Indiana University-Purdue University, Indianapolis, IN, USA 46202; \\
\supit{b}Department of Physics, Arizona State University, Tempe, AZ, USA 46202; \\
\supit{c}Department of Biomedical Sciences, Shahid Beheshti Medical School, Tehran, IRAN 
}

\begin{document} 
  \maketitle 

%%%%%%%%%%%%%%%%%%%%%%%%%%%%%%%%%%%%%%%%%%%%%%%%%%%%%%%%%%%%% 
To appear in: Proceedings Volume 11318, Medical Imaging 2020: Imaging Informatics for Healthcare, Research, and Applications; 113180E (2020) https://doi.org/10.1117/12.2548526
Event: SPIE Medical Imaging, 2020, Houston, Texas, United States

\begin{abstract}
Imaging methods by using computer techniques provide doctors assistance at any time and relieve their workload, especially for iterative processes like identifying objects of interest such as lesions and anatomical structures from the image. Decetion of microaneurysms (MAs) as a one of the lesions in the retina is considered to be a crucial step in some retinal image analysis algorithms for identification of diabetic retinopathy (DR) as the second largest eye diseases in developed countries. The objective of this study is to compare effect of two preprocessing methods, Illumination Equalization, and Top-hat transformation, on retinal images to detect MAs using combination of Matching based approach and deep learning methods either in the normal fundus images or in the presence of DR. The steps for the detection are as following: 1) applying preprocessing, 2) vessel segmentation and masking, and 3) MAs detection using combination of Matching based approach and deep learning. From the accuracy view point, we compared the method to manual detection performed by ophthalmologists for our big retinal image databases (more than 2200 images). Using first preprocessing method, Illumination equalization and contrast enhancement, the accuracy of MAs detection was about 90\% for all databases (one local and two publicly retinal databases). The performance of the MAs detection methods using top-hat preprocessing (the second preprocessing method) was more than 80\% for all databases.
\end{abstract}

%>>>> Include a list of keywords after the abstract 

%\keywords{Manuscript format, template, SPIE Proceedings, LaTeX}

%%%%%%%%%%%%%%%%%%%%%%%%%%%%%%%%%%%%%%%%%%%%%%%%%%%%%%%%%%%%%
\section{INTRODUCTION}
\label{sec:intro}  % \label{} allows reference to this section

\begin{figure}
   \begin{center}
   \begin{tabular}{c}
   \includegraphics[height=6cm]{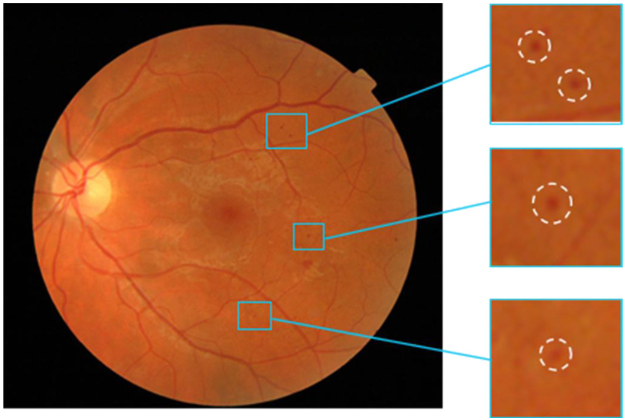}
   \end{tabular}
   \end{center}
   \caption[example] 
%>>>> use \label inside caption to get Fig. number with \ref{}
   { \label{fig:MAsamples} 
A sample retinal image from local database with different size of MAs.}
   \end{figure}

The transparent living tissue of eye causes the retina to be the only part of body where vascular network is directly visible. Many systemic diseases change the retinal vessel networks and could be diagnosed through this transparent window. However, the secure check of retina to find abnormality is doing by ophthalmologists, which is elaborating task and associated with consuming time, error and fatigue. Moreover, clinical analysis is based on doctor's idea which may not be repeatable. One possible solution for these issues is to use Computer Assisted Diagnosis (CAD) systems. Regarding eye diseases, one of the causes of creating abnormalities in the retina is coming from Diabetic retinopathy (DR)~\cite{tavakoli2010early,  pourreza2009automatic}. DR is becoming a major public health issue because it is one of the main sources of vision lost and blindness~\cite{lee2015epidemiology, tavakoli2017automated-onh}. Microaneurysms (MAs), which are small bobble appearing on the side of tiny retinal blood vessels, are the most frequent and the first lesions to appear as a sign of DR~\cite{niemeijer2009retinopathy, tavakoli2013complementary, abramoff2008evaluation}. 
Fig.~\ref{fig:MAsamples} shows a retinal image with multiple MAs with different diameters usually ranges from $10\mu m$ to $120 \mu m$~\cite{pereira2014using, walter2007automatic}, which are considerably less than the size of other retinal landmarks such as optic disk.
%-------------
%-------------

One of big difficulties in detection of MAs is detection of them is done manually by the ophthalmologist and in the same direction the number of people who have DR and afflicted by MAs are continue for growing at an alarming rate~\cite{thomas2015prevalence, pedro2010prevalence}. However, only 50\% of them are aware of DR. 
Therefore, seting up automated DR detection systems has received a lot of attention from the research community. The computer methods are used for providing eye doctors assistance at any time and to relieve their workload or iterative works as well, to identify object of interest such as lesions and anatomical structures from the image~\cite{tavakoli2017automated-fovea, tavakoli2017automated-onh}. 
Detection of MAs is the critical step in the automated detection of DR. They are visible directly after the arterial phase of fluorescein angiography ~\cite{tavakoli2013complementary, walter2007automatic}. In the same direction, numbering of MAs has been used as a tool for evaluation of the progress of the DR~\cite{tavakoli2013complementary}.
The goal of this study is detecting these type of lesions. The MAs have a number of characteristics that can be used in their detection: (1) Gaussian shape of the MA cross-sectional grey level profile~\cite{lazar2012retinal}. (2) The MAs are circular~\cite{tavakoli2013complementary}, (3) Their size  which is less than $125 \mu m$~\cite{walter2007automatic}. Before MAs detection in fundus image, the image has to be preprocessed to ensure adequate level of success in detection. Here we applied two different preprocessing methods (Illumination equalization-contrast enhancement, and top-hat transform)~\cite{tavakoli2017comparing, tavakoli2017effect} separately and compare the results of MAs detection in each of these two ways. After preprocessing we used Laplacian of Gaussian (LoG) edge detector for retinal vessel segmentation~\cite{tavakoli2017automated-onh, tavakoli2017automated-fovea}.
After vessel segmentation and masking, a hybrid approach, combining Matching based method and deep learning, was proposed for detection of all the MAs. 

There are several studies for automated detection of MAs in color retinal images. 
These appraoches can be generally classified into three different methods including morphological operation, and template matching, as unsupervised methods~\cite{lazar2012retinal, giancardo2010microaneurysms, tavakoli2013complementary}, and supervised learning~\cite{tavakoli2013complementary, niemeijer2005automatic, dashtbozorg2018retinal, habib2017detection, walter2007automatic, quellec2008optimal, wang2016localizing, zhang2010detection, gegundez2017tool, ram2010successive, lazar2012retinal, mizutani2009automated, dai2018clinical}.
The main benefit of unsupervised approaches is that they do not need a training phase~\cite{habib2017incorporating}. 
Some unsupervised MA detection methods that have been presented are Gaussian filters~\cite{streeter2003microaneurysm, fleming2006automated, niemeijer2005automatic} or their variants~\cite{zhang2010detection, wu2015new}, simple thresholding~\cite{tavakoli2013complementary, giancardo2010microaneurysms}, double ring filter~\cite{mizutani2009automated}, mixture model-based clustering~\cite{sanchez2009mixture} one dimession lines scanning~\cite{lazar2012retinal}, Hessian matrix Eigenvalues~\cite{adal2014automated}, and Frangi filters~\cite{srivastava2015red}.
Different adjustment was applied based on morphology approach to increase the detection accuracy~\cite{fleming2006automated, mizutani2009automated}. Although this type of processing typically is fast and easy to apply, the ability of the approach is limited by the its builder. In better words, some main hidden structures and uncover patterns could be ignored by the builder and cause false segmentation~\cite{dai2018clinical}. 
Several other mathematical morphology based methods proposed for the detection of red lesions.
According to statistical result, the intensity distribution of MAs is matched to Gaussian distribution~\cite{quellec2008optimal, zhang2010detection, lazar2012retinal}. Therefore, template matching based MA detection approaches were proposed and greatly improved the detection accuracy. 

By growing of machine learning ideas~\cite{tavakoli2019pitching, tavakoli2019bayesian}, studies on MA detection using classification based approaches are mostly seen recently. 
A variety of classification approaches have been presented such as Linear Discriminant Analysis (LDA)~\cite{streeter2003microaneurysm} K- Nearest Neighbours (KNN)~\cite{niemeijer2005automatic, fleming2006automated, adal2014automated}, Artificial Neural Networks~\cite{mizutani2009automated, rosas2015method}, Naive Bayes~\cite{sopharak2013simple} and Logistic Regression~\cite{garcia2010assessment}.
Antal et al.~\cite{antal2012ensemble} applied a rule-based expert system for MAs detection. In this approach, after selection of MA candidates from retinal images, a rule-based classifier is applied to find true MAs.
Niemeijer et al.~\cite{niemeijer2005automatic} used a hybrid strategy using both top-hat based approach  and a supervised classification. In this approach, MA candidates same as  Antal et al.~\cite{antal2012ensemble} were first selected and then a classifier was trained to differentiate true MAs from false ones. although these machine learning based approaches succeed in detecting hidden structures of features and MAs, they still rely on manually selected features and empirically determined parameters~\cite{dai2018clinical}.
Furthermore, related to third category of detection, learning based approaches,~\cite{gulshan2016development, zhou2017automatic, gargeya2017automated, seoud2015red, abramoff2016improved, orlando2018ensemble, chudzik2018microaneurysm, dai2018clinical, costa2018weakly} proposed to address above issues. 
Gulshan et al.~\cite{gulshan2016development} presented a deep learning based algorithm to automatically grade DR in retinal images. In this method, deep neural network is applied to process directly the images and output the grading result of DR. While this work successfully addressed the problem of finding hidden structures and empirically determined parameters is not need, for training purpose the classifiers it requires large amounts of retinal images and their annotations, which is costly and time-consuming. Moreover, there is not any  quantitative results generated explicitly for a specific MA. In fact for understanding of the development of DR and monitoring its progress, these quantitative data are critical.
Seoud et al.~\cite{seoud2015red} proposed a novel method for automatic detection of both MAs in color retinal images. The main focus of their work is a new set of shape features, called Dynamic Shape Features, that do not need to precisely segment of red lesion regions. The approach detects all types of MAs  in the images, without distinguishing between them. Differentiating between these lesion types is really important in the clinical practice. The detection of MA is not a practical solution in the medical field. 
Haloi~\cite{haloi2015improved} implemented five layers deep learning with drop out mechanism for diagnosing of early stage DR. The shortcome of the approach was the requirement for a large amount of training data and time-consuming~\cite{wu2017automatic}. 
In comparison, our work focuses on detecting and distinguishing MAs in fundus images. Here, before working on post processing step which we are using the concept of deep learning we add preprocessing unsupervised steps to have some candidates as the MAs and among them using the deep learning we are looking for final true MAs.
%%%%%%%%%%%%%%%%%%%%%%%%%%%%%%%%%%%%%%%%%%%%%%%%%%%%%%%%%%%%%
\section{Methods} 

%\subsection{Materials} 
%\label{sec:Materials}

%%-----------------------------------------------------------
\subsection{Preprocessing} 
\label{sec:title}

The preprocessing step provides us a retinal image with clear lesions, and landmarks. Moreover, preprocessed image provide higher contrast between the retinal objects and background which unifies the histogram of the images. Here we used two different preprocessing methods separately and compare the results of MAs detection in each of these two ways:

\subsubsection{Illumination Equalization, and Contrast Enhancement}
\begin{figure}
   \begin{center}
   \begin{tabular}{c}
   \includegraphics[height=5cm]{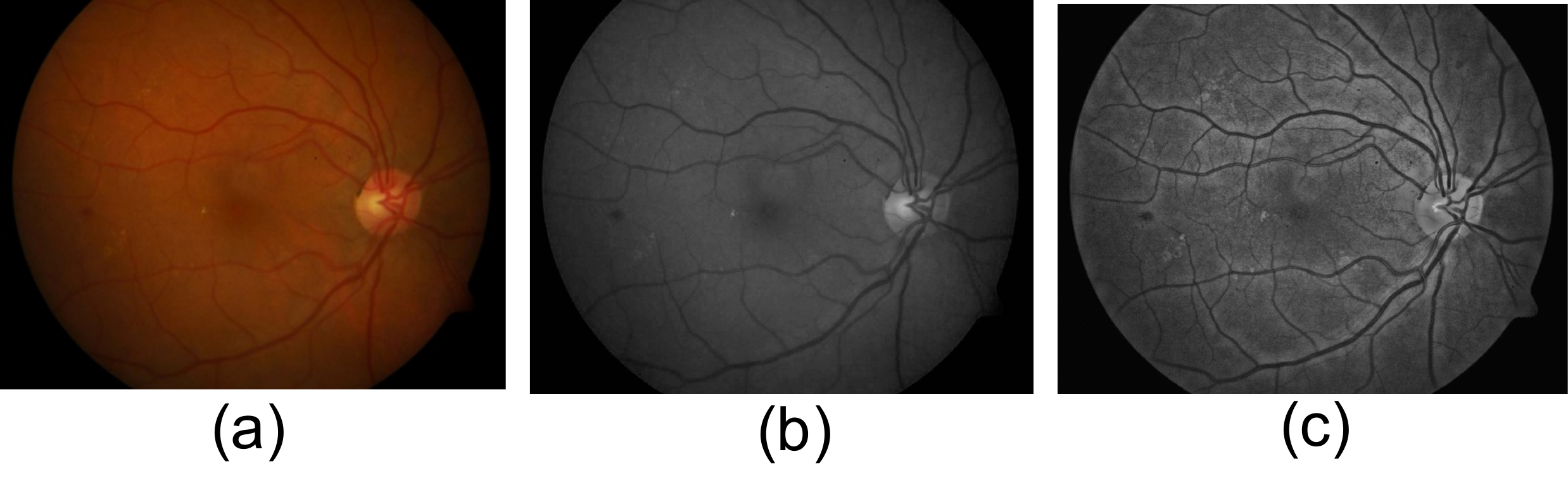}
   \end{tabular}
   \end{center}
   \caption[example] 
%>>>> use \label inside caption to get Fig. number with \ref{}
   { \label{fig:Illumination-Equalization} 
Original RGB image from local dataset (a), original G channel (b), its preprocessing results using Illumination Equalization, and Contrast Enhancement (c)}
   \end{figure}

In Illumination equalization and contrast enhancement, our purpose is to unify the histogram of all fundus images. In background elimination, background of the image is filtered to some level and we process just the foreground image. For this preprocessing level, there are three steps that described in this study~\cite{tavakoli2017effect, tavakoli2017comparing}: color space conversion, illumination equalization, and contrast enhancement. At the beginning, the RGB retinal image is converted to the HSV color system. In the next step, we search for retinal region detection, and to do this we use the region of the retina from the HSV image. In the third step, the Illumination equalization, the original RGB image takes as an input and equalizes uneven illumination in the image. The output of this part represents the channels of the original image where the illumination has been equalized. At the end, in contrast enhancement we select a reference image and use its color histogram as template for all other images to normalize the background. In better words, since each fundus image has its own brightness, this approach helps us to use the same threshold for all images. For this purpose, a reference image is chosen and the histogram is adjusted. Then histogram specification is used to incorporate the images. As we have shown in Fig.~\ref{fig:Illumination-Equalization}, by performing preprocessing, we have a retinal image with maximum possible contrast between the retinal landmarks  and background and also unify the histogram of the available fundus images.

%-------------
%-------------

\subsubsection{Top-hat transformation}
\begin{figure}
   \begin{center}
   \begin{tabular}{c}
   \includegraphics[height=5cm]{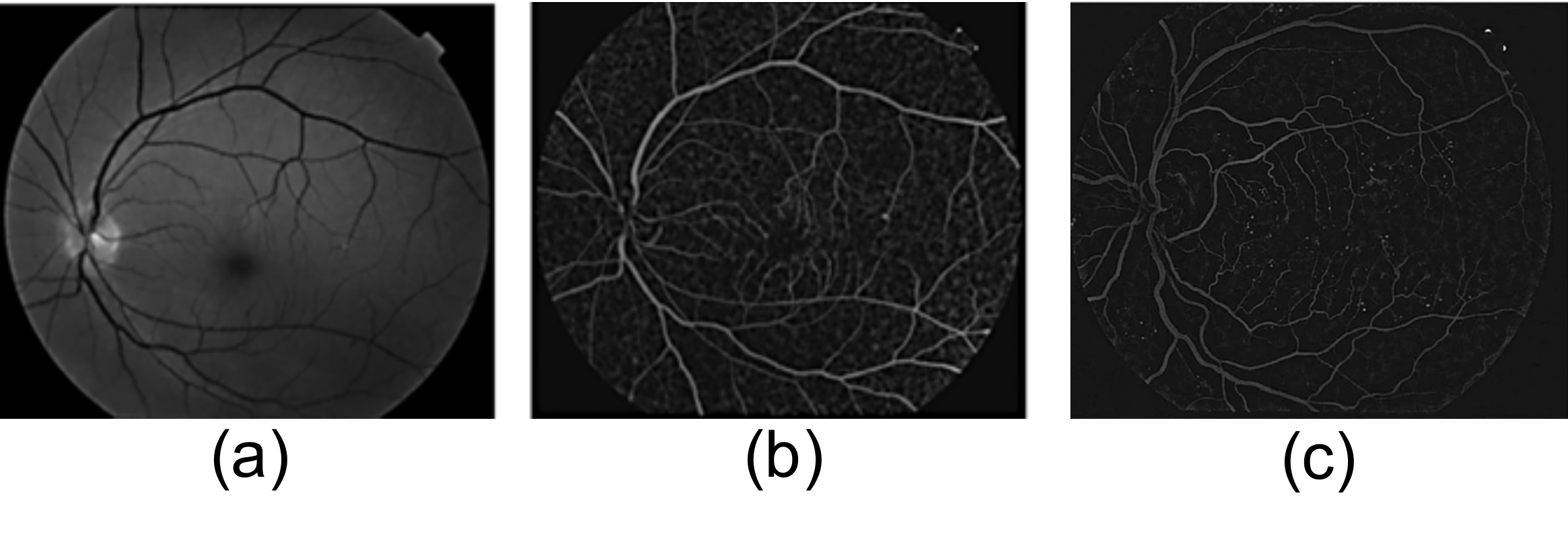}
   \end{tabular}
   \end{center}
   \caption[example] 
%>>>> use \label inside caption to get Fig. number with \ref{}
   { \label{fig:top-hat} 
Preprocessing green channel image from local database (a), using top-hat preprocessing (b), applying averaging filter on top-hat result (c)}
   \end{figure}

Here among the three components (R, G, B) of retinal images, we use green channel that has the best contrast between vessel and background; so the green channel is selected as input image. As the second preprocessing step, the top-hat transform is one of the important morphological operators which we apply here~\cite{tavakoli2017effect, tavakoli2017comparing, tavakoli2013complementary}. In the top-hat the basic idea is increasing the contrast between the landmarks (vessels), lesions (MAs) with the  background regions of the image. This transformation extracts bright and dim image regions corresponding to the applied structure element. Our top-hat transformation was based on a disk structure element whose diameter was empirically found that the best compromise between the retinal features of interest and background. The disk diameter depended on the input image resolution. After applying top-hat, we use contrast stretching to change the brightness of an image. The output image was a linear mapping of a subset of pixel values to the entire range of grays, from the black to the white, producing an image with much higher contrast. Filtering a region is the process of applying a filter to a region of interest in an image, where a binary mask defines the region. In this situation we used an averaging filter on the result of image from last section (top-hat result) after that, we subtract the image from last section with the result of applying averaging filter. Before this section, in image result from top-hat transform, there are some variations in the image and some points like noise that without eliminate these points maybe supposed as some part of vessels, that increase false positive rate of our algorithm and after applying filter and subtraction this variation was removed. In other words, this section is applied for better removing background variation for better detecting of vessels. The result of first step is shown in Fig.~\ref{fig:top-hat}.

%-------------
%-------------

%%-----------------------------------------------------------
\subsection{Vessel Segmentation and Masking} 
In this study, the steps for the vessel detection are in following: (1) Smoothing: suppress as much noise as possible, without destroying the true edges, (2) Enhancement: apply a filter to enhance the quality of the edges in the image (sharpening), (3) Detection: determine which edge pixels should be discarded as noise and which should be retained by thresholding the edge strength and edge size, (4) Localization: determine the exact location of an edge by edge thinning or linking. To do vessel segmentation we used Laplacian-of-Gaussian (LoG)~\cite{tavakoli2017automated-onh, tavakoli2017comparing}. In the LoG edge detector uses the second-order spatial differentiation: $\bigtriangledown^2f= \frac{\partial^2 f}{\partial x^2}+\frac{\partial^2 f}{\partial y^2}$

The Laplacian is usually combined with smoothing as a precursor to finding edges via zero-crossings. The 2-D Gaussian function: $h(x,y)=e^{\frac{-(x^2 + y^2)}{2\sigma^2}}$. Where $\sigma$ is the standard deviation, blurs the image with the degree of blurring being determined by the value of $\sigma$. If an image is pre-smoothed by a Gaussian filter, then we have the LoG operation that is defined: $\bigtriangledown^2 G_{\sigma}\ast I$ where $\bigtriangledown^2 G_{\sigma}(x,y) = \frac{1}{2\pi \sigma^4}(\frac{x^2 + y^2}{\sigma^2} - 2)e^{\frac{-(x^2 + y^2)}{2\sigma^2}}$ (See Fig.~\ref{fig:MA})

%%-----------------------------------------------------------
\subsection{Microaneurysm Detection} 
To remove all interfering effects, we masked vessel tree in the retinal image. In this section to extract circular patterns a Matching Based approach following with concept of deep learning and CNNs was a way which not only was utilized for MAs detection also simplified the statistical analysis of the input retinal image. 
To detect MAs, circular patterns with diameter lower than $125 \mu m$~\cite{walter2007automatic} should be excerpted in local sub-images (windows). Therefore, the maximum size of window was chosen twice more than size of the biggest MA. 
The size of our Matching filter was selected equal to maximum diameter of the biggest MA in pixel. Here we found that 18 pixels for MUMS-DB, and 10 pixels for DRIVE database empirically. On the other hand, point noise, end point of vessels, and bifurcations are similar to MAs (false MAs). Therefore, to validate MAs, following MAs characteristics in the images were used: Intensity, Size, and Shape. Thresholding is a way to evaluate the intensity. In other words, an easy solution to the MAs validation problem is to compare the peak amplitude intensity with predefined thresholds. Size and shape of candidate were checked.
Unlike unsupervised methods which has no initial labels and must find natural clustering patterns in the data, deep learning approach is a learning model that can be applied for classification and regression analysis. Deep learning takes in the clustered bag of features and their corresponding labels (MAs or nonMAs) and determines the predictor clusters for each class. The process is to first train the hidden layers which is done by MA and nonMA candidates. We used CNN implemented in MATLAB~\cite{vedaldi2015matconvnet}. The purpose of the CNN classification layer is simply to transform all the net activations to a series of values that can be interpreted as probabilities in the final output layer. To do this, the CNN
MATLAB toolbox is applied onto the net outputs. %By using appropriate preprocessing and also accurate vascular detection method, automatic system could mask some parts of vessels that were assumed MAs (false MAs). (See Fig. 3)

A total number of 2240 color fundus images were labeled independently with sufficient quality by an expert ophthalmologists with more than 15 years experience in diagnosing DR at early stages. Match filtering produces the MA candidates. These retinal images were made into sub-images, centered on finding the MAs. 25000 sub-images
were used which mutually agreed with the accuracy
of their clinical label. Among these, 70\% were used
for the training purpose and remaining in the testing set.
Sub-images were extracted at twice the size of the biggest
MA ($125 \mu m$). Normal sub-images (or sub-images without
MA) were taken from the same image as the abnormal
sub-images (or sub-images with MA) in regions that were
free of MA. Overall, 15000 sub-images containing MA and
the remaining sub-images without MA were taken from the
whole sub-images.

\begin{figure}
   \begin{center}
   \begin{tabular}{c}
   \includegraphics[height=8cm]{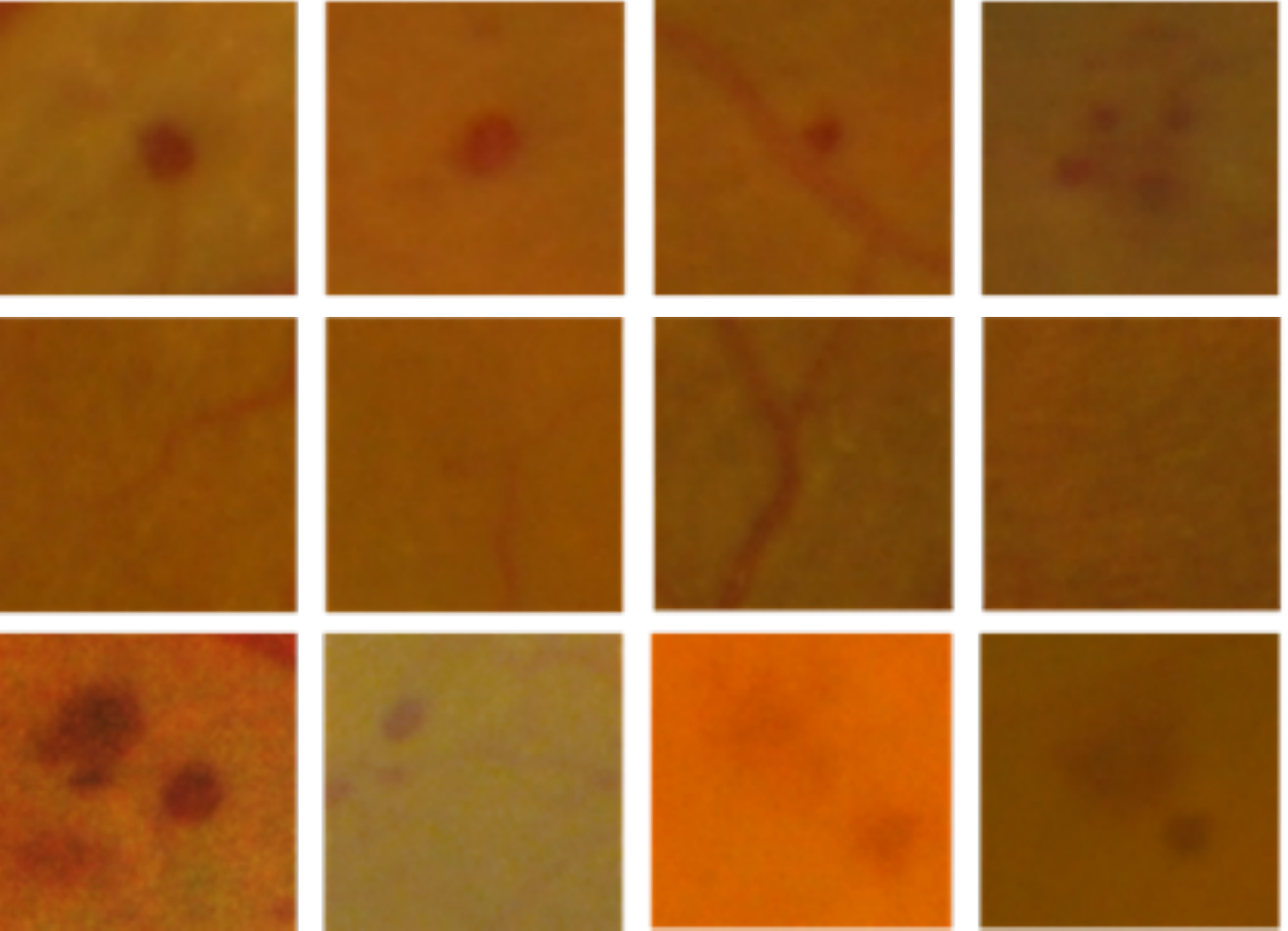}
   \end{tabular}
   \end{center}
   \caption[example] 
%>>>> use \label inside caption to get Fig. number with \ref{}
   { \label{fig:CNN-Candids} 
Examples of sub-images containing candidates of interest.
First row different MAs; Second row end
points of vessels and bifurcation points, Last row combination of MAs, and vessels.}
   \end{figure}

A single CNN using the MATLAB architecture with a $128 \times 128 \times 3$ input and a four-class output was designed: (1) normal, (2) MAs, (3) bifurcation points, and (4) end points of retinal vessels. Some samples from the different classes are shown in Fig. \ref{fig:CNN-Candids}. The CNN was trained on 17500 sub-images (70\%) and tested on 7500 selected sub-images. Training and testing were performed using the MATLAB Deep Learning toolbox \cite{vedaldi2015matconvnet}.

%-------------
%-------------

\begin{figure}
   \begin{center}
   \begin{tabular}{c}
   \includegraphics[height=12cm]{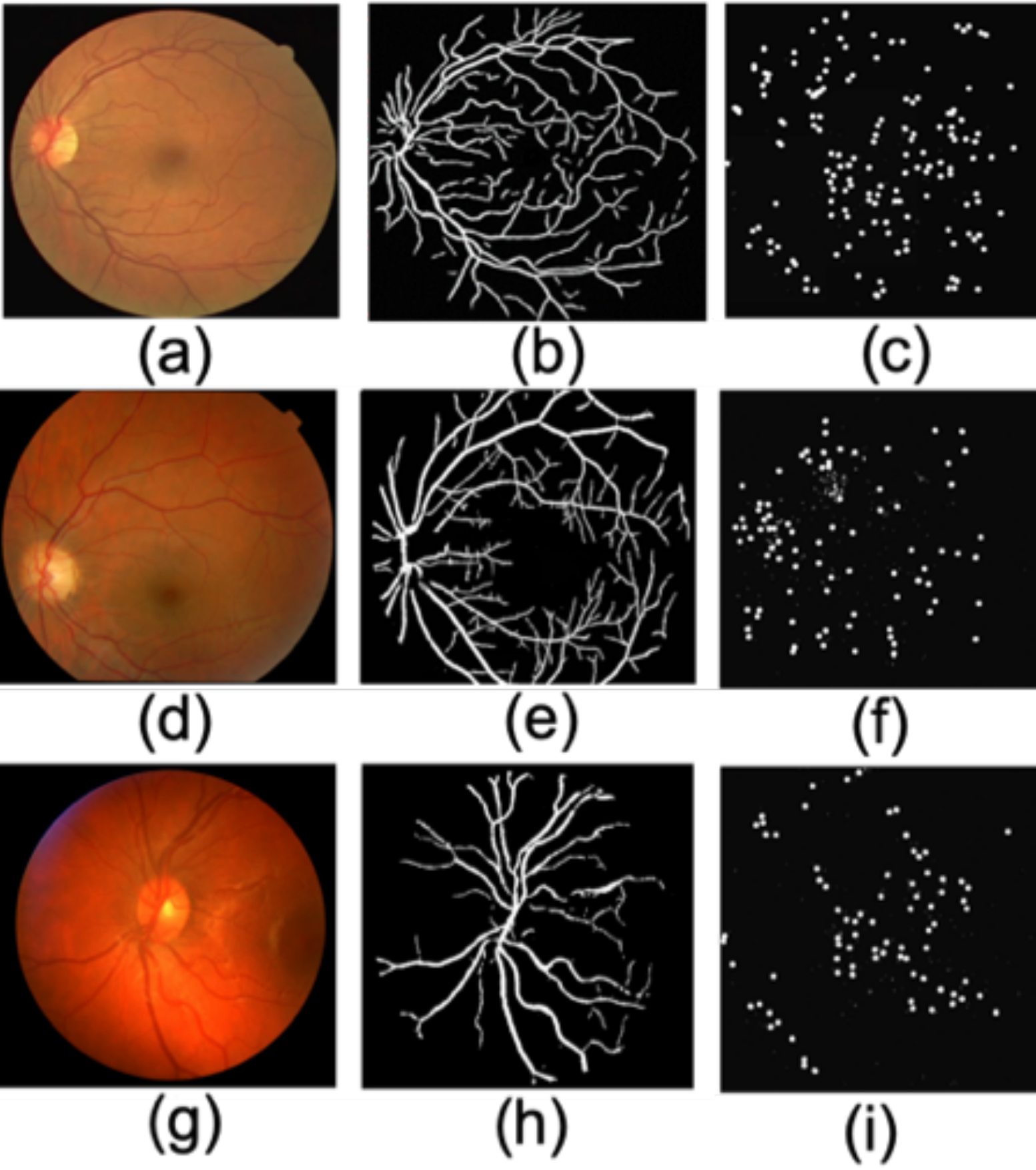}
   \end{tabular}
   \end{center}
   \caption[example] 
%>>>> use \label inside caption to get Fig. number with \ref{}
   { \label{fig:MA} 
Results of vessel segmentation and masking and MAs detection for all three databases. (a-c) DRIVE (first public dataset), (d-f) MUMS-DB (local dataset), (g-i) MESSIDOR (second public dataset)}
   \end{figure}

As we mentioned, we used preprocessed images until now. Using an overlapping sliding window, the trained CNN was employed over the full scan of the image. A $20 \times 20 \times 3$ window was moved across full sub-images with a slide of 5 pixels overlapping. Each window was the input for a forward pass through our trained CNN, and produce a probability score within that sub-image for each of the four classes of normal and MAs. The result of this sliding window was a blanket of probability values over the entire image for each of the four classes. This procedure took about 1.5 minutes using a PC desktop with an Intel i5-5600HQ Processor. 
The results of MA detection has been shown in the Fig.~\ref{fig:MA}.

%-------------
%-------------

%%-----------------------------------------------------------
\section{Results} 
In this section, statistical information about the sensitivity and specificity measures is extracted. The higher the sensitivity and specificity values, the better the procedure. 
The results for the automated method compared to the groundtruth or gold standard were calculated for each image. These metrics are defined as:

\begin{equation} \label{eq:sensitivity}
\begin{aligned}
Sensitivity  &= \frac{TP}{TP+FN} \\
Specificity  &= \frac{TN}{TN+FP} \\
\end{aligned}
\end{equation}

Where TP is true positive, TN is true negative, FP is false positive and FN is false negative same as \cite{tavakoli2013complementary, marin, tavakoli2019quantitative}.

For the MAs detection, three datasets, one rural and two publicly available databases, were used. The first rural database was named Mashhad University Medical Science Database (MUMS-DB). A total of 1000 color images were captured which 920 were with DR and 80 normal from both left and right eyes of patients. Images were taken with a field of view (FOV) of 50 degrees. The acquired image resolution is $2896 \times 1944$ TIFF format~\cite{tavakoli2011automated, pourreza2014computationally, tavakoli2011radon}. The second database, DRIVE, consisting of 40 images with image resolution of $768 \times 584$ pixels in which 33 cases did not have any sign of DR and 7 ones showed signs of early or mild DR with a 45 FOV~\cite{niemeijer2004comparative}. The last database, MESSIDOR includes 1200 eye fundus color images of the posterior pole with a 45 degree FOV. The images were acquired using 8 bits per color plane at three differnt resolution  $1440\times960$, $2240\times1488$, and  $2304\times1536$ pixels~\cite{decenciere2014feedback}

Statistical information about the sensitivity and specificity measures is extracted. For all retinal images of test set (2240 images for the test purpose and training), our reader labeled the MAs on the images and the result of this manual segmentation are saved to be analyzed further. By using first preprocessing method, Illumination equalization, and contrast enhancement, the sensitivity and specificity of MAs detection was about 90\% and 85\% for all databases. The performance of the MAs detection methods using top-hat preprocessing (the second preprocessing method) was 80\% for both sensitivity and specificity for all databases.
Reaching to the sensitivity of more than 80\% makes our CAD system as good as or better than other related published studies~\cite{quellec2008optimal, zhang2010detection, lazar2012retinal, habib2017detection, wu2015new}. Moreover, this sensitivity in MA based analysis show the ability of our algorithm even in treatment planning and follows up~\cite{orlando2018ensemble}. However, the disadvantage of most of the proposed approaches was that it did not avoid the overfitting issue and was not able to introduce a standard feature selection principle. Bisides, for instance, in~\cite{zhang2010detection} they ignored the hidden and unnoticeable structures. Moreover, a lot of parameters need empirically to be determined.

%%-----------------------------------------------------------
\section{Conclusion} 
In this study we were focused only on detection of MAs. The goal of this work was to develop an algorithm for detection of vascular lesions, MAs, related to DR. %and development of an automated detection of DR. 
Since these days medical images are in the digital format, it is feasible to establish a computer-based system that automatically detects landmarks from these images images~\cite{tavakoli2019quantitative, tavakoli2019quantitative-spect}. An automated screening system would save the working load of medical doctors, and letting clinics to use their assets in other important tasks. It could also be doable to check more people and more often by assisting of an automatic screening system, since it would be more inexpensive than screening by humans~\cite{tavakoli2016single}. 
Moreover, computers are suitable to issues involving the derivation of quantitative information from images because of their capacity to process data in fast and efficient manner with a high degree of reproducibility~\cite{matsopoulos1999automatic, tavakoli2017attenuation, welikala2015genetic}.
In this paper, we proposed combination of matching based approach and deep learning to detect all MAs from color fundus retinal image.
The results proved that it is possible to use our CAD system for assisting an ophthalmologist to segment fundus images into normal parts and lesions, and thus support the ophthalmologist in his or her decision making. To utilize this program in the follow up of patients, we should add an image registration algorithm so that the ophthalmologist could study the effect of his/her treatment and also the progression of the disease, not only by crisp counting, but also by spatial orientation which is included in presented method. The presented approach was evaluated through a public retinal image databases DRIVE and MESSIDOR. The experiment results demonstrated that using the first preprocessing approach combine with matched based method and deep learning  has better detection performance in terms of of sensitivity in comparison with other published studies~\cite{quellec2008optimal, lazar2012retinal, habib2017detection, wu2015new}.

%%%%%%%%%%%%%%%%%%%%%%%%%%%%%%%%%%%%%%%%%%%%%%%%%%%%%%%%%%%%%

%%%%%%%%%%%%%%%%%%%%%%%%%%%%%%%%%%%%%%%%%%%%%%%%%%%%%%%%%%%%%
%%%%% References %%%%%

\bibliography{report}   %>>>> bibliography data in report.bib

\begin{thebibliography}{10}

\bibitem{tavakoli2010early}
M.~Tavakoli, A.~Mehdizadeh, R.~Pourreza, T.~Banaee, M.~H. Bahreyni~Toossi, and
  H.~R. Pourreza, ``Early detection of diabetic retinopathy in fluorescent
  angiography retinal images using image processing methods,'' {\em Iranian
  Journal of Medical Physics}~{\bf 7}(4), pp.~7--14, 2010.

\bibitem{pourreza2009automatic}
H.~R. Pourreza, M.~H. Bahreyni~Toossi, A.~Mehdizadeh, R.~Pourreza, and
  M.~Tavakoli, ``Automatic detection of microaneurysms in color fundus images
  using a local radon transform method,'' {\em Iranian Journal of Medical
  Physics}~{\bf 6}(1), pp.~13--20, 2009.

\bibitem{lee2015epidemiology}
R.~Lee, T.~Y. Wong, and C.~Sabanayagam, ``Epidemiology of diabetic retinopathy,
  diabetic macular edema and related vision loss,'' {\em Eye and vision}~{\bf
  2}(1), p.~17, 2015.

\bibitem{tavakoli2017automated-onh}
M.~Tavakoli, M.~Nazar, A.~Golestaneh, and F.~Kalantari, ``Automated optic nerve
  head detection based on different retinal vasculature segmentation methods
  and mathematical morphology,'' in {\em 2017 IEEE Nuclear Science Symposium
  and Medical Imaging Conference (NSS/MIC)},  pp.~1--7, IEEE, 2017.

\bibitem{niemeijer2009retinopathy}
M.~Niemeijer, B.~Van~Ginneken, M.~J. Cree, A.~Mizutani, G.~Quellec, C.~I.
  S{\'a}nchez, B.~Zhang, R.~Hornero, M.~Lamard, C.~Muramatsu, {\em et~al.},
  ``Retinopathy online challenge: automatic detection of microaneurysms in
  digital color fundus photographs,'' {\em IEEE transactions on medical
  imaging}~{\bf 29}(1), pp.~185--195, 2009.

\bibitem{tavakoli2013complementary}
M.~Tavakoli, R.~P. Shahri, H.~Pourreza, A.~Mehdizadeh, T.~Banaee, and M.~H.~B.
  Toosi, ``A complementary method for automated detection of microaneurysms in
  fluorescein angiography fundus images to assess diabetic retinopathy,'' {\em
  Pattern Recognition}~{\bf 46}(10), pp.~2740--2753, 2013.

\bibitem{abramoff2008evaluation}
M.~D. Abr{\`a}moff, M.~Niemeijer, M.~S. Suttorp-Schulten, M.~A. Viergever,
  S.~R. Russell, and B.~Van~Ginneken, ``Evaluation of a system for automatic
  detection of diabetic retinopathy from color fundus photographs in a large
  population of patients with diabetes,'' {\em Diabetes care}~{\bf 31}(2),
  pp.~193--198, 2008.

\bibitem{pereira2014using}
C.~Pereira, D.~Veiga, J.~Mahdjoub, Z.~Guessoum, L.~Gon{\c{c}}alves,
  M.~Ferreira, and J.~Monteiro, ``Using a multi-agent system approach for
  microaneurysm detection in fundus images,'' {\em Artificial intelligence in
  medicine}~{\bf 60}(3), pp.~179--188, 2014.

\bibitem{walter2007automatic}
T.~Walter, P.~Massin, A.~Erginay, R.~Ordonez, C.~Jeulin, and J.-C. Klein,
  ``Automatic detection of microaneurysms in color fundus images,'' {\em
  Medical image analysis}~{\bf 11}(6), pp.~555--566, 2007.

\bibitem{thomas2015prevalence}
R.~L. Thomas, F.~D. Dunstan, S.~D. Luzio, S.~R. Chowdhury, R.~V. North, S.~L.
  Hale, R.~L. Gibbins, and D.~R. Owens, ``Prevalence of diabetic retinopathy
  within a national diabetic retinopathy screening service,'' {\em British
  Journal of Ophthalmology}~{\bf 99}(1), pp.~64--68, 2015.

\bibitem{pedro2010prevalence}
R.-A. Pedro, S.-A. Ramon, B.-B. Marc, F.-B. Juan, and M.-M. Isabel,
  ``Prevalence and relationship between diabetic retinopathy and nephropathy,
  and its risk factors in the north-east of spain, a population-based study,''
  {\em Ophthalmic epidemiology}~{\bf 17}(4), pp.~251--265, 2010.

\bibitem{tavakoli2017automated-fovea}
M.~Tavakoli, P.~Kelley, M.~Nazar, and F.~Kalantari, ``Automated fovea detection
  based on unsupervised retinal vessel segmentation method,'' in {\em 2017 IEEE
  Nuclear Science Symposium and Medical Imaging Conference (NSS/MIC)},
  pp.~1--7, IEEE, 2017.

\bibitem{lazar2012retinal}
I.~Lazar and A.~Hajdu, ``Retinal microaneurysm detection through local rotating
  cross-section profile analysis,'' {\em IEEE transactions on medical
  imaging}~{\bf 32}(2), pp.~400--407, 2012.

\bibitem{tavakoli2017comparing}
M.~Tavakoli, F.~Kalantari, and A.~Golestaneh, ``Comparing different
  preprocessing methods in automated segmentation of retinal vasculature,'' in
  {\em 2017 IEEE Nuclear Science Symposium and Medical Imaging Conference
  (NSS/MIC)},  pp.~1--8, IEEE, 2017.

\bibitem{tavakoli2017effect}
M.~Tavakoli, M.~Nazar, and A.~Mehdizadeh, ``Effect of two different
  preprocessing steps in detection of optic nerve head in fundus images,'' in
  {\em Medical Imaging 2017: Computer-Aided Diagnosis},   {\bf 10134},
  p.~101343A, International Society for Optics and Photonics, 2017.

\bibitem{giancardo2010microaneurysms}
L.~Giancardo, F.~M{\'e}riaudeau, T.~P. Karnowski, K.~W. Tobin, Y.~Li, and
  E.~Chaum, ``Microaneurysms detection with the radon cliff operator in retinal
  fundus images,'' in {\em Medical Imaging 2010: Image Processing},   {\bf
  7623}, p.~76230U, International Society for Optics and Photonics, 2010.

\bibitem{niemeijer2005automatic}
M.~Niemeijer, B.~Van~Ginneken, J.~Staal, M.~S. Suttorp-Schulten, and M.~D.
  Abr{\`a}moff, ``Automatic detection of red lesions in digital color fundus
  photographs,'' {\em IEEE Transactions on medical imaging}~{\bf 24}(5),
  pp.~584--592, 2005.

\bibitem{dashtbozorg2018retinal}
B.~Dashtbozorg, J.~Zhang, F.~Huang, and B.~M. ter Haar~Romeny, ``Retinal
  microaneurysms detection using local convergence index features,'' {\em IEEE
  Transactions on Image Processing}~{\bf 27}(7), pp.~3300--3315, 2018.

\bibitem{habib2017detection}
M.~Habib, R.~Welikala, A.~Hoppe, C.~Owen, A.~Rudnicka, and S.~Barman,
  ``Detection of microaneurysms in retinal images using an ensemble
  classifier,'' {\em Informatics in Medicine Unlocked}~{\bf 9}, pp.~44--57,
  2017.

\bibitem{quellec2008optimal}
G.~Quellec, M.~Lamard, P.~M. Josselin, G.~Cazuguel, B.~Cochener, and C.~Roux,
  ``Optimal wavelet transform for the detection of microaneurysms in retina
  photographs,'' {\em IEEE transactions on medical imaging}~{\bf 27}(9),
  pp.~1230--1241, 2008.

\bibitem{wang2016localizing}
S.~Wang, H.~L. Tang, Y.~Hu, S.~Sanei, G.~M. Saleh, T.~Peto, {\em et~al.},
  ``Localizing microaneurysms in fundus images through singular spectrum
  analysis,'' {\em IEEE Transactions on Biomedical Engineering}~{\bf 64}(5),
  pp.~990--1002, 2016.

\bibitem{zhang2010detection}
B.~Zhang, X.~Wu, J.~You, Q.~Li, and F.~Karray, ``Detection of microaneurysms
  using multi-scale correlation coefficients,'' {\em Pattern Recognition}~{\bf
  43}(6), pp.~2237--2248, 2010.

\bibitem{gegundez2017tool}
M.~E. Gegundez-Arias, D.~Marin, B.~Ponte, F.~Alvarez, J.~Garrido, C.~Ortega,
  M.~J. Vasallo, and J.~M. Bravo, ``A tool for automated diabetic retinopathy
  pre-screening based on retinal image computer analysis,'' {\em Computers in
  biology and medicine}~{\bf 88}, pp.~100--109, 2017.

\bibitem{ram2010successive}
K.~Ram, G.~D. Joshi, and J.~Sivaswamy, ``A successive clutter-rejection-based
  approach for early detection of diabetic retinopathy,'' {\em IEEE
  Transactions on Biomedical Engineering}~{\bf 58}(3), pp.~664--673, 2010.

\bibitem{mizutani2009automated}
A.~Mizutani, C.~Muramatsu, Y.~Hatanaka, S.~Suemori, T.~Hara, and H.~Fujita,
  ``Automated microaneurysm detection method based on double ring filter in
  retinal fundus images,'' in {\em Medical Imaging 2009: Computer-Aided
  Diagnosis},   {\bf 7260}, p.~72601N, International Society for Optics and
  Photonics, 2009.

\bibitem{dai2018clinical}
L.~Dai, R.~Fang, H.~Li, X.~Hou, B.~Sheng, Q.~Wu, and W.~Jia, ``Clinical report
  guided retinal microaneurysm detection with multi-sieving deep learning,''
  {\em IEEE transactions on medical imaging}~{\bf 37}(5), pp.~1149--1161, 2018.

\bibitem{habib2017incorporating}
M.~Habib, R.~Welikala, A.~Hoppe, C.~G. Owen, A.~R. Rudnicka, A.~Tufail,
  C.~Egan, and S.~A. Barman, ``Incorporating spatial information for
  microaneurysm detection in retinal images,'' {\em Advances in Science,
  Technology and Engineering Systems Journal}~{\bf 2}(3), pp.~642--649, 2017.

\bibitem{streeter2003microaneurysm}
L.~Streeter and M.~J. Cree, ``Microaneurysm detection in colour fundus
  images,'' {\em Image Vision Comput. New Zealand} , pp.~280--284, 2003.

\bibitem{fleming2006automated}
A.~D. Fleming, S.~Philip, K.~A. Goatman, J.~A. Olson, and P.~F. Sharp,
  ``Automated assessment of diabetic retinal image quality based on clarity and
  field definition,'' {\em Investigative ophthalmology \& visual science}~{\bf
  47}(3), pp.~1120--1125, 2006.

\bibitem{wu2015new}
J.~Wu, J.~Xin, L.~Hong, J.~You, and N.~Zheng, ``New hierarchical approach for
  microaneurysms detection with matched filter and machine learning,'' in {\em
  2015 37th Annual International Conference of the IEEE Engineering in Medicine
  and Biology Society (EMBC)},  pp.~4322--4325, IEEE, 2015.

\bibitem{sanchez2009mixture}
C.~I. S{\'a}nchez, R.~Hornero, A.~Mayo, and M.~Garc{\'\i}a, ``Mixture
  model-based clustering and logistic regression for automatic detection of
  microaneurysms in retinal images,'' in {\em Medical Imaging 2009:
  Computer-Aided Diagnosis},   {\bf 7260}, p.~72601M, International Society for
  Optics and Photonics, 2009.

\bibitem{adal2014automated}
K.~M. Adal, D.~Sidib{\'e}, S.~Ali, E.~Chaum, T.~P. Karnowski, and
  F.~M{\'e}riaudeau, ``Automated detection of microaneurysms using
  scale-adapted blob analysis and semi-supervised learning,'' {\em Computer
  methods and programs in biomedicine}~{\bf 114}(1), pp.~1--10, 2014.

\bibitem{srivastava2015red}
R.~Srivastava, D.~W. Wong, L.~Duan, J.~Liu, and T.~Y. Wong, ``Red lesion
  detection in retinal fundus images using frangi-based filters,'' in {\em 2015
  37th Annual International Conference of the IEEE Engineering in Medicine and
  Biology Society (EMBC)},  pp.~5663--5666, IEEE, 2015.

\bibitem{tavakoli2019pitching}
M.~Tavakoli, S.~Jazani, I.~Sgouralis, O.~M. Shafraz, B.~Donaphon,
  S.~Sivasankar, M.~Levitus, and S.~Presse, ``Pitching single focus confocal
  data analysis one photon at a time with bayesian nonparametrics,'' {\em
  bioRxiv} , p.~749739, 2019.

\bibitem{tavakoli2019bayesian}
M.~Tavakoli, S.~Jazani, I.~Sgouralis, and S.~Presse, ``Bayesian nonparametrics
  for fluorescence methods,'' {\em Biophysical Journal}~{\bf 116}(3), p.~39a,
  2019.

\bibitem{rosas2015method}
R.~Rosas-Romero, J.~Mart{\'\i}nez-Carballido, J.~Hern{\'a}ndez-Capistr{\'a}n,
  and L.~J. Uribe-Valencia, ``A method to assist in the diagnosis of early
  diabetic retinopathy: Image processing applied to detection of microaneurysms
  in fundus images,'' {\em Computerized medical imaging and graphics}~{\bf 44},
  pp.~41--53, 2015.

\bibitem{sopharak2013simple}
A.~Sopharak, B.~Uyyanonvara, and S.~Barman, ``Simple hybrid method for fine
  microaneurysm detection from non-dilated diabetic retinopathy retinal
  images,'' {\em Computerized Medical Imaging and Graphics}~{\bf 37}(5-6),
  pp.~394--402, 2013.

\bibitem{garcia2010assessment}
M.~Garc{\'\i}a, M.~I. L{\'o}pez, D.~{\'A}lvarez, and R.~Hornero, ``Assessment
  of four neural network based classifiers to automatically detect red lesions
  in retinal images,'' {\em Medical engineering \& physics}~{\bf 32}(10),
  pp.~1085--1093, 2010.

\bibitem{antal2012ensemble}
B.~Antal and A.~Hajdu, ``An ensemble-based system for microaneurysm detection
  and diabetic retinopathy grading,'' {\em IEEE transactions on biomedical
  engineering}~{\bf 59}(6), pp.~1720--1726, 2012.

\bibitem{gulshan2016development}
V.~Gulshan, L.~Peng, M.~Coram, M.~C. Stumpe, D.~Wu, A.~Narayanaswamy,
  S.~Venugopalan, K.~Widner, T.~Madams, J.~Cuadros, {\em et~al.}, ``Development
  and validation of a deep learning algorithm for detection of diabetic
  retinopathy in retinal fundus photographs,'' {\em Jama}~{\bf 316}(22),
  pp.~2402--2410, 2016.

\bibitem{zhou2017automatic}
W.~Zhou, C.~Wu, D.~Chen, Y.~Yi, and W.~Du, ``Automatic microaneurysm detection
  using the sparse principal component analysis-based unsupervised
  classification method,'' {\em IEEE Access}~{\bf 5}, pp.~2563--2572, 2017.

\bibitem{gargeya2017automated}
R.~Gargeya and T.~Leng, ``Automated identification of diabetic retinopathy
  using deep learning,'' {\em Ophthalmology}~{\bf 124}(7), pp.~962--969, 2017.

\bibitem{seoud2015red}
L.~Seoud, T.~Hurtut, J.~Chelbi, F.~Cheriet, and J.~P. Langlois, ``Red lesion
  detection using dynamic shape features for diabetic retinopathy screening,''
  {\em IEEE transactions on medical imaging}~{\bf 35}(4), pp.~1116--1126, 2015.

\bibitem{abramoff2016improved}
M.~D. Abr{\`a}moff, Y.~Lou, A.~Erginay, W.~Clarida, R.~Amelon, J.~C. Folk, and
  M.~Niemeijer, ``Improved automated detection of diabetic retinopathy on a
  publicly available dataset through integration of deep learning,'' {\em
  Investigative ophthalmology \& visual science}~{\bf 57}(13), pp.~5200--5206,
  2016.

\bibitem{orlando2018ensemble}
J.~I. Orlando, E.~Prokofyeva, M.~del Fresno, and M.~B. Blaschko, ``An ensemble
  deep learning based approach for red lesion detection in fundus images,''
  {\em Computer methods and programs in biomedicine}~{\bf 153}, pp.~115--127,
  2018.

\bibitem{chudzik2018microaneurysm}
P.~Chudzik, S.~Majumdar, F.~Caliv{\'a}, B.~Al-Diri, and A.~Hunter,
  ``Microaneurysm detection using fully convolutional neural networks,'' {\em
  Computer methods and programs in biomedicine}~{\bf 158}, pp.~185--192, 2018.

\bibitem{costa2018weakly}
P.~Costa, A.~Galdran, A.~Smailagic, and A.~Campilho, ``A weakly-supervised
  framework for interpretable diabetic retinopathy detection on retinal
  images,'' {\em IEEE Access}~{\bf 6}, pp.~18747--18758, 2018.

\bibitem{haloi2015improved}
M.~Haloi, ``Improved microaneurysm detection using deep neural networks,'' {\em
  arXiv preprint arXiv:1505.04424} , 2015.

\bibitem{wu2017automatic}
B.~Wu, W.~Zhu, F.~Shi, S.~Zhu, and X.~Chen, ``Automatic detection of
  microaneurysms in retinal fundus images,'' {\em Computerized Medical Imaging
  and Graphics}~{\bf 55}, pp.~106--112, 2017.

\bibitem{vedaldi2015matconvnet}
A.~Vedaldi and K.~Lenc, ``Matconvnet: Convolutional neural networks for
  matlab,'' in {\em Proceedings of the 23rd ACM international conference on
  Multimedia},  pp.~689--692, ACM, 2015.

\bibitem{marin}
D.~Mar{\'\i}n, A.~Aquino, M.~E. Geg{\'u}ndez-Arias, and J.~M. Bravo, ``A new
  supervised method for blood vessel segmentation in retinal images by using
  gray-level and moment invariants-based features,'' {\em IEEE Transactions on
  Medical Imaging}~{\bf 30}(1), pp.~146--158, 2011.

\bibitem{tavakoli2019quantitative}
M.~Tavakoli, K.~Tsekouras, R.~Day, K.~W. Dunn, and S.~Press{\'e},
  ``Quantitative kinetic models from intravital microscopy: A case study using
  hepatic transport,'' {\em The Journal of Physical Chemistry B}~{\bf 123}(34),
  pp.~7302--7312, 2019.

\bibitem{tavakoli2011automated}
M.~Tavakoli, M.~B. Toosi, R.~Pourreza, T.~Banaee, and H.~R. Pourreza,
  ``Automated optic nerve head detection in fluorescein angiography fundus
  images,'' in {\em 2011 IEEE Nuclear Science Symposium Conference Record},
  pp.~3057--3060, IEEE, 2011.

\bibitem{pourreza2014computationally}
R.~Pourreza-Shahri, M.~Tavakoli, and N.~Kehtarnavaz, ``Computationally
  efficient optic nerve head detection in retinal fundus images,'' {\em
  Biomedical Signal Processing and Control}~{\bf 11}, pp.~63--73, 2014.

\bibitem{tavakoli2011radon}
M.~Tavakoli, A.~Mehdizadeh, R.~Pourreza, H.~R. Pourreza, T.~Banaee, and M.~B.
  Toosi, ``Radon transform technique for linear structures detection:
  application to vessel detection in fluorescein angiography fundus images,''
  in {\em 2011 IEEE Nuclear Science Symposium Conference Record},
  pp.~3051--3056, IEEE, 2011.

\bibitem{niemeijer2004comparative}
M.~Niemeijer, J.~Staal, B.~van Ginneken, M.~Loog, and M.~D. Abramoff,
  ``Comparative study of retinal vessel segmentation methods on a new publicly
  available database,'' in {\em Medical imaging 2004: image processing},   {\bf
  5370}, pp.~648--657, International Society for Optics and Photonics, 2004.

\bibitem{decenciere2014feedback}
E.~Decenci{\`e}re, X.~Zhang, G.~Cazuguel, B.~Lay, B.~Cochener, C.~Trone,
  P.~Gain, R.~Ordonez, P.~Massin, A.~Erginay, {\em et~al.}, ``Feedback on a
  publicly distributed image database: the messidor database,'' {\em Image
  Analysis \& Stereology}~{\bf 33}(3), pp.~231--234, 2014.

\bibitem{tavakoli2019quantitative-spect}
M.~Tavakoli and M.~Neij, ``Quantitative evaluation of the effect of attenuation
  correction in {SPECT} images with ct-derived attenuation,'' in {\em Medical
  Imaging 2019: Physics of Medical Imaging},   {\bf 10948}, p.~109485U,
  International Society for Optics and Photonics, 2019.

\bibitem{tavakoli2016single}
M.~Tavakoli, J.~N. Taylor, C.-B. Li, T.~Komatsuzaki, and S.~Press{\'e},
  ``Single molecule data analysis: An introduction,'' {\em arXiv preprint
  arXiv:1606.00403} , 2016.

\bibitem{matsopoulos1999automatic}
G.~K. Matsopoulos, N.~A. Mouravliansky, K.~K. Delibasis, and K.~S. Nikita,
  ``Automatic retinal image registration scheme using global optimization
  techniques,'' {\em IEEE Transactions on Information Technology in
  Biomedicine}~{\bf 3}(1), pp.~47--60, 1999.

\bibitem{tavakoli2017attenuation}
M.~Tavakoli, M.~Naji, A.~Abdollahi, and F.~Kalantari, ``Attenuation correction
  in {SPECT} images using attenuation map estimation with its emission data,''
  in {\em Medical Imaging 2017: Physics of Medical Imaging},   {\bf 10132},
  p.~101324Z, International Society for Optics and Photonics, 2017.

\bibitem{welikala2015genetic}
R.~A. Welikala, M.~M. Fraz, J.~Dehmeshki, A.~Hoppe, V.~Tah, S.~Mann, T.~H.
  Williamson, and S.~A. Barman, ``Genetic algorithm based feature selection
  combined with dual classification for the automated detection of
  proliferative diabetic retinopathy,'' {\em Computerized Medical Imaging and
  Graphics}~{\bf 43}, pp.~64--77, 2015.

\end{thebibliography}
\bibliographystyle{spiebib}   %>>>> makes bibtex use spiebib.bst

\end{document}